# Dynamic and structural properties of orthorhombic rare-earth manganites under high pressure

**D. A. Mota[1], A. Almeida[1], V. H. Rodrigues[2], M. M. R. Costa[2], P. Tavares[3], P. Bouvier[4], M. Guennou[5], J. Kreisel[5,6], J. Agostinho Moreira[1]**

*1-IFIMUP and IN-Institute of Nanoscience and Nanotechnology, Departamento de Física e Astronomia da Faculdade de Ciências, Universidade do Porto, Rua do Campo Alegre, 687, 4169-007 Porto, Portugal.*

*2-CEMDRX, Departamento de Física, Faculdade de Ciências e Tecnologia, Universidade de Coimbra, 3004-516 Coimbra, Portugal.*

*3-Centro de Química - Vila Real, Departamento de Química. Universidade de Trás-os-Montes e Alto Douro, 5000-801 Vila Real, Portugal.*

*4-Laboratoire des Matériaux et du Génie Physique, CNRS, Université Grenoble-Alpes, 3 parvis Louis Néel, 38016 Grenoble, France.*

*5-Département Science et Analyse des Matériaux, CRP Gabriel Lippmann, 41 rue du Brill, L-4422 Belvaux, Luxembourg.*

*6-Physics and Materials Science Research Unit, University of Luxembourg, 41 Rue du Brill, L-4422 Belvaux, Luxembourg*

jamoreir@fc.up.pt

## Abstract

We report a high-pressure study of orthorhombic rare-earth manganites $A\mathrm{MnO_3}$ using Raman scattering (for *A* = Pr, Nd, Sm, Eu, Tb and Dy) and synchrotron X-ray diffraction (for *A* = Pr, Sm, Eu, and Dy). In all cases, a structural and insulator-to-metal transition was evidenced, with a critical pressure that depends on the *A*-cation size. We analyze the compression mechanisms at work in the different manganites via the pressure dependence of the lattice parameters, the shear strain in the *a*-*c* plane, and the Raman bands associated with out-of-phase $\mathrm{MnO_6}$ rotations and in-plane O2 symmetric stretching modes. Our data show a crossover across the rare-earth series between two different kinds of behavior. For the smallest *A*-cations, the compression is nearly isotropic in the *ac* plane, with presumably only very slight changes of tilt angles and Jahn-Teller distortion. As the radius of the *A*-cation increases, the pressure-induced reduction of Jahn-Teller distortion becomes more pronounced and increasingly significant as a compression mechanism, while the pressure-induced bending of octahedra chains becomes conversely less pronounced. We finally discuss our results in the light of the notion of chemical pressure, and show that the analogy with hydrostatic pressure works quite well for manganites with small A-cations but can be misleading with large *A*-cations.

## 1. Introduction

The transition metal oxides with perovskite structure have attracted a huge interest due to their remarkable functional properties: ferroelectricity, superconductivity, magnetoresistivity, magnetism, multiferroicity, among others [1]. This versatility finds its origin in the couplings and competitions between different phenomena with similar energies (high polarizability of the crystal lattice, and strong electronic correlations). This gives rise to different phases whose stabilization strongly depends on external parameters like temperature, magnetic/electric field, chemical substitution, strain and hydrostatic pressure. Among the transition metal oxides, the rare-earth manganites, $A\mathrm{MnO_3}$, have been extensively studied. These compounds crystallize at ambient conditions into the P*nma* space group [2]. The main distortions in orthomanganites are the rotations (tilts) of the $MnO_6$ octahedra with $a^-b^+a^-$ tilting scheme in Glazer's notation, and a cooperative Jahn-Teller distortion with orbital ordering. The structural distortions are intimately linked to the physical properties. It is notably well established that the balance between the competitive antiferromagnetic and ferromagnetic interactions in rare-earth manganites can be changed by varying the tilt of the $MnO_6$ octahedra, pointing out for a strong spin-lattice coupling [3,4]. Much work has therefore been devoted to the fine tuning of these distortions by cation substitution or external parameters.

Among the possible external parameters, hydrostatic pressure has this specificity that it allows modifying the interatomic distances and, thus the interactions, to a much larger extent than any other parameter like temperature, strain or magnetic field. However, high-pressure studies on manganites are still relatively scarce. The emphasis has been mostly laid on the model compound $LaMnO_3$. A first report by Loa *et al.* [5] of the suppression of the Jahn-Teller distortion at 18 GPa followed by an insulator-to-metal transition at 32 GPa triggered a number of theoretical studies devoted to Jahn-Teller distortion under pressure and the mechanism responsible for the insulator-to-metal transition [6-9]. Experimentally, Raman and EXAFS studies have given evidence for the persistence of the Jahn-Teller distortion over the entire range of the insulating phase at the local scale [10,11]. In contrast, only a couple of works have been reported on other manganites. Combined x-ray diffraction and x-ray absorption spectroscopy studies on $TbMnO_3$ and $DyMnO_3$ showed a gradual decrease of the Jahn-Teller distortion and a bandwidth broadening, but could not clarify the presence of insulator-to-metal transition itself [12,13]. High-pressure Raman study of $TbMnO_3$ and $PrMnO_3$ up to 10 GPa [14] were focused on the compression mechanisms as inferred from the pressure evolution of Raman frequencies. Recently, we reported synchrotron x-ray diffraction and Raman spectroscopy studies of $GdMnO_3$ under pressure up to 55 GPa and showed evidence for a first-order structural transition accompanied by an insulator-to-metal transition around 50 GPa [15]. In contrast, Muthu *et al.* have not observed any structural transition at all up to 50 GPa for *A* = Eu, Gd, Tb and Dy [16].

In spite of these efforts, a systematic overview of the effect of hydrostatic pressure on rare-earth manganites is still lacking, and open questions remain. One central point is the persistence or disappearance of the Jahn-Teller distortion at high pressure, both on the average macroscopic scale and on the local scale. This point is strongly linked to the possible mechanisms for the insulator-to-metal transition at high pressure, whose very presence needs to be checked for a number of compounds. In addition, it is unclear whether or not this insulator-to-metal phase transition, when it exists, is accompanied by a symmetry-breaking structural phase transition.

The aim of this work is to contribute to this overview by a study of rare-earth manganites in the 50-60 GPa range using Raman spectroscopy and x-ray powder diffraction (XRD). We investigate the effect of the *A*-cation radius on the compression behavior, the insulator-to-metal and structural transition and the evolution of octahedra tilts and Jahn-Teller distortion. The paper is structured as follows: after experimental details (section 2), we first describe the structure and Raman spectra of the different compounds at ambient conditions (section 3.a) before moving the high-pressure results in Raman spectroscopy (section 3.b) and XRD (section 3.c). Different aspects are discussed in section 4: high-pressure phase transitions (4.a), compression mechanisms (4.b), spontaneous strains and Jahn-Teller distortion (4.c) and finally chemical pressure (4.d).

## 2. Experimental details

High quality ceramic $A$MnO$_3$ ($A$ = Pr, Nd, Sm, Eu, Tb and Dy) samples were prepared with the sol-gel urea combustion method [17] and its chemical, morphological and structural characteristics were checked by powder XRD, SEM, EDS and XPS. All high-pressure experiments were carried out at room temperature. The powder samples of $A$MnO$_3$ were loaded in a Diamond Anvil Cell (DAC) with diamond cullets of 300 µm diameter and with Helium as a pressure-transmitting medium. The Raman spectra were recorded on a LabRam spectrometer using a He-Ne laser at 633 nm. The laser power was kept below 5 mW on the DAC to avoid sample heating. The spectra were fitted with a sum of independent damped oscillators. High pressure synchrotron x-ray powder diffraction experiments were performed on the Extreme Conditions Beamline (ECB) at PETRA III-Desy, and at the ESRF on the ID27 high pressure beamline. XRD patterns were collected on a Perkin Elmer image detector at wavelength λ = 0.29118 Å for $PrMnO_3$ and $SmMnO_3$ (PETRAIII) and on a Mar CCD detector at λ = 0.3738 Å for $EuMnO_3$ and $DyMnO_3$ (ID27 - ESRF). Diffraction patterns were obtained after integration of raw data using FIT2D [18]. The powder diffraction data were analyzed by Le Bail refinements using the FullProf software [19]. The refined parameters were the lattice parameters, the parameters of a pseudo-Voigt profile and the zero shift. The background was fitted by linear interpolation between a set of manually chosen points for each diffractogram. The coordinates of these points were not refined.

## 3. Experimental Results

### a. Raman scattering at ambient conditions. General Considerations.

At ambient conditions, the studied rare-earth manganites crystallize in an orthorhombic structure with space group P*nma* with four formula units per unit cell. The Raman activity in these compounds arises from structural deviations to the ideal cubic perovskite symmetry. The basic distortions involved in symmetry reduction from P*m-3m* to P*nma* are the [010] and [101] rotations of the $MnO_6$ octahedra. Moreover, as the $Mn^{3+}$ ion is Jahn-Teller active, a cooperative Jahn-Teller distortion (CJTD) occurs, but does not lower the symmetry any further. Due to octahedra tilting and CJTD, the rare-earth ion ($A$-cation) is displaced from the high-symmetry position (at the center of the eight surrounding $MnO_6$ octahedra) and thus its coordination reduces from 12 to 8, but it remains in the mirror plane. The octahedral rotations cause the equatorial long Mn-O bonds to align more closely with the $a$ axis than with the $c$ axis. Due to these distortions, the four oxygen atoms belonging to the equatorial plane of the $MnO_6$ octahedra (here after designated by O2) occupy general positions, while the $Mn^{3+}$ ion remains on the inversion center and the apical oxygen atoms (designated by O1) remain in the mirror plane. Factor group analysis provides the following decomposition corresponding to the 60 normal vibrations at the Γ-point of the Brillouin zone:

$$\Gamma_{acoustic} = B_{1u}+B_{2u}+B_{3u}$$

$$\Gamma_{optical} = (7A_g+5B_{1g}+7B_{2g}+5B_{3g})_{Raman\text{-}active} + (8A_u+9B_{1u}+7B_{2u}+9B_{3u})_{IR\text{-}active}.$$

Note that the vibrations involving $Mn^{3+}$ do not contribute to the Raman spectra because $Mn^{3+}$ sits on an inversion center and Raman-active modes preserve the inversion center.

Figure 1 shows the unpolarized Raman spectra of *A*$MnO_3$, with *A* = Pr, Nd, Sm, Eu, Gd, Tb and Dy, recorded at ambient conditions. Due to the polycrystalline nature of the studied samples, our Raman spectra exhibit simultaneously all Raman-active modes $A_g$, $B_{1g}$, $B_{2g}$ and $B_{3g}$. The Raman peaks are more clearly separated as the *A*-cation size decreases, due to the corresponding increase of the orthorhombic distortion. Our Raman spectra are in excellent agreement with those previously reported in the literature [20,21] and we shall use throughout the paper the mode assignment defined in Ref. 21. In the following, we shall focus our attention on the Raman bands whose activation is associated with the main structural deformations allowing for symmetry reduction as these bands will be further used to monitor the pressure-induced deformations in the *A*$MnO_3$ compounds.

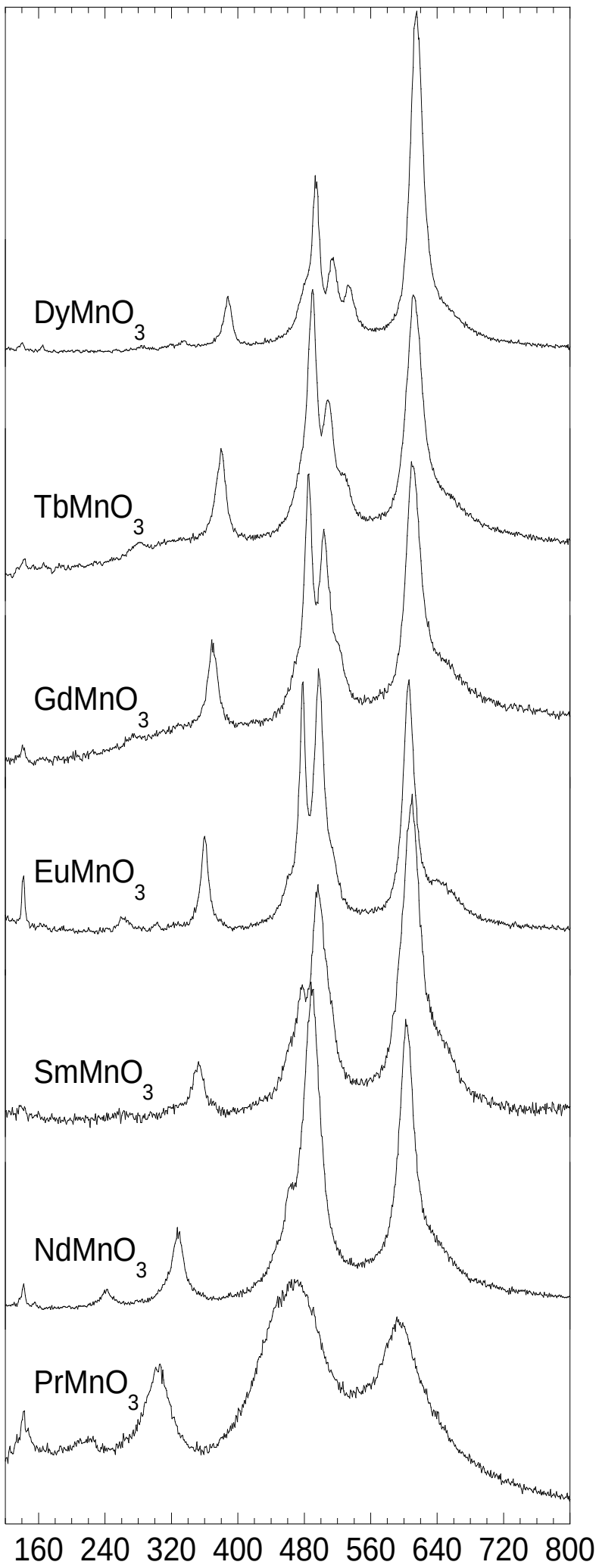

**Figure 1**. Unpolarized Raman spectra of the *A*$MnO_3$ compounds at ambient conditions.

In Figure 2(a) and (b), we present the wave number of the Raman bands as a function of the ionic radius of the rare-earth ions with coordination VIII, assigned to the in-plane O2 symmetric stretching modes (labelled $B_{2g}(1)$ following Ref. 21), and the out-of-phase $MnO_6$ rotations (Ag(4)), respectively. The out-of-phase $MnO_6$ rotation, with $A_g$ symmetry, is activated by the [101] rotations [21], while the in-plane O2 symmetric stretching mode, with $B_{2g}$ symmetry, is activated by the Jahn-Teller distortion.

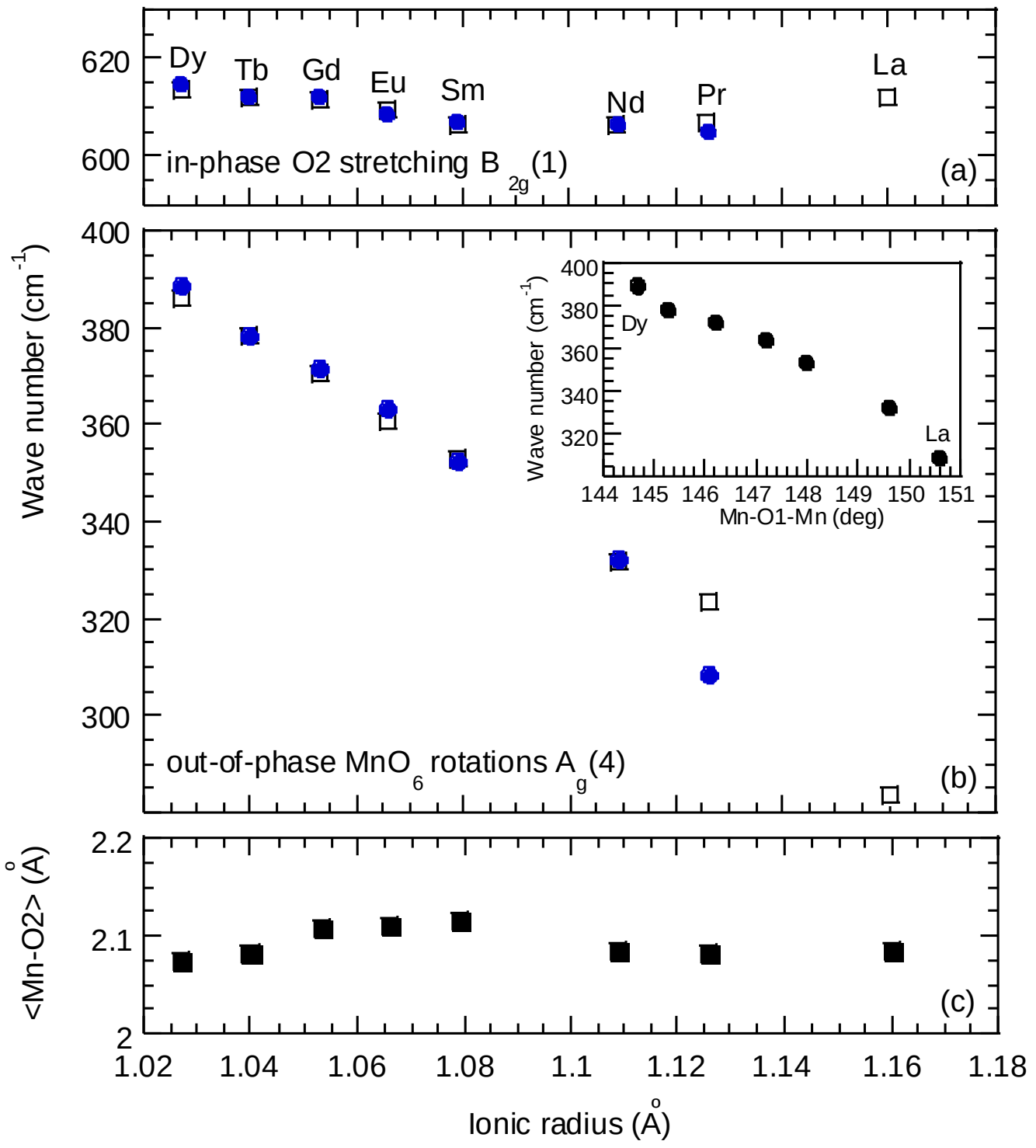

**Figure 2**. Evolution of (a) the wavenumber of the in-plane O2 symmetric stretching mode, (b) the wavenumber of the out-of-phase $MnO_6$ rotations, and (c) the mean Mn-O2 bond length, as a function of the ionic radius (in coordination VIII). Our data (closed symbols) are compared to the data from Ref. 21 (open symbols). The inset of Figure 2(b) shows the wave number of the out-of-phase $MnO_6$ rotations as a function of the Mn-O1-Mn bond angle.

The inset of Figure 2(b) shows the wave number of the tilt mode as a function of the Mn-O1-Mn bond angle for the studied rare-earth manganites, obtained from the refinement of the XRD patterns obtained at room conditions [22-25]. As already shown in manganites and other perovskites, the tilt angle scales with the wavenumber of the tilt mode [21,26]. On the other hand, the in-plane O2 symmetric stretching mode (see Figure 2(a)), hereafter designated by symmetric stretching mode, exhibits a more complex dependence on the rare-earth ionic radius. This mode involves stretching vibrations of the equatorial plane O2 atoms, whose wave number is determined by the Mn-O2 distances. The mean Mn-O2 bond length value is plotted in Figure 2(c) as a function of the ionic radius (coordination VIII). As the ionic radius decreases from *A* = La to Eu, the mean Mn-O2 bond length increases and the wave number decreases, reaching its minimum value for $SmMnO_3$. Then, on further ionic radius decreasing, the mean Mn-O2 bond length decreases and the value of the wave number of the symmetric stretching mode increases. Moreover, both wave number of the symmetric stretching mode and mean Mn-O2 bond length do not vary significantly with the rare-earth ionic radius, so we can conclude that the Mn-O2 bond lengths exhibit a weak dependence on volume reduction due to the decrease of the A-site size. From the results presented up to this point, we can conclude that the volume reduction driven by *A*-site ionic radius decrease is mainly accommodated by bending the $MnO_6$ network.

### b. High pressure Raman spectroscopy

Figure 3 displays representative Raman spectra recorded at selected hydrostatic pressures.

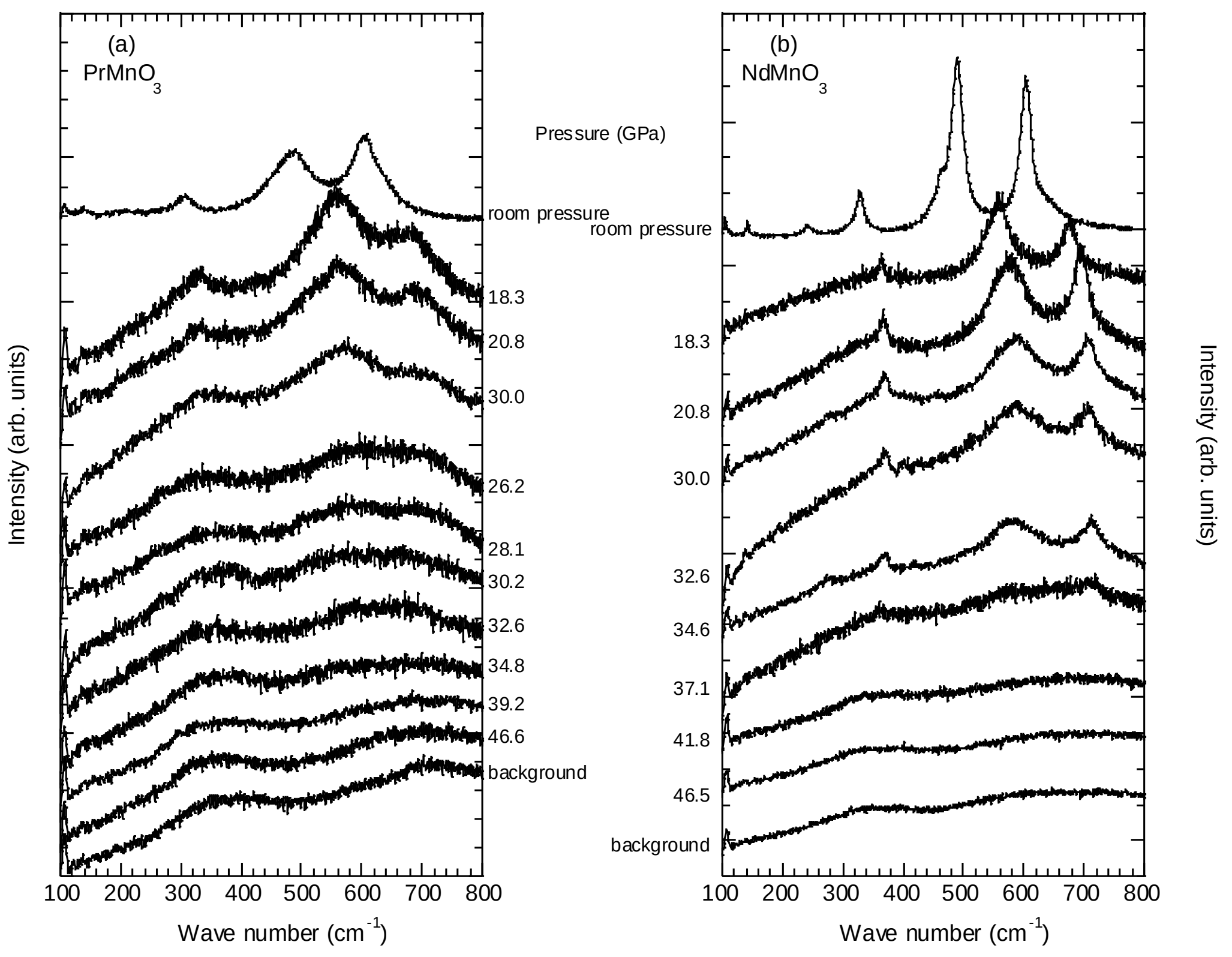

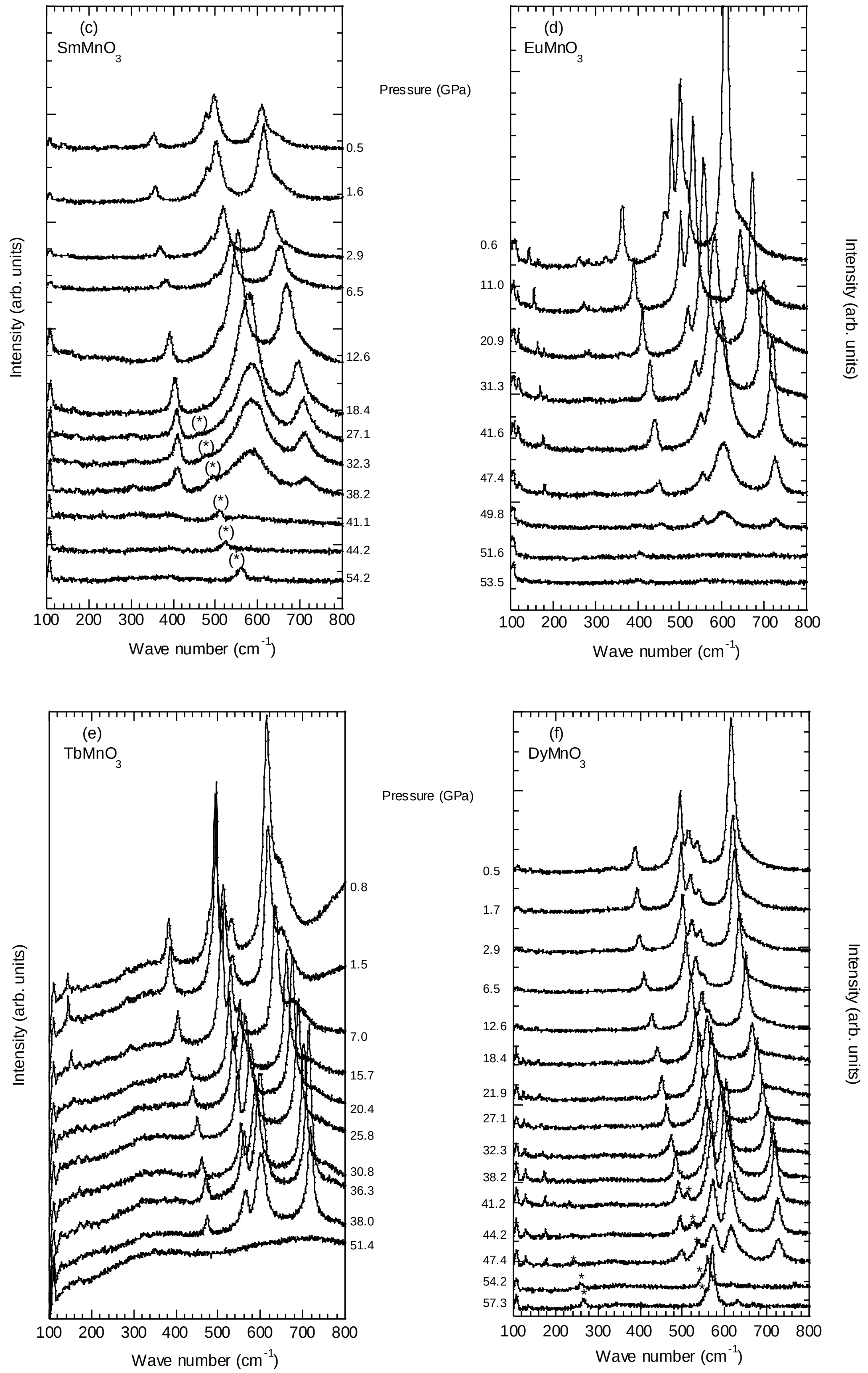

**Figure 3.** Raman spectra of the *A*$MnO_3$ compounds, recorded at room temperature and at fixed hydrostatic pressure. The bands marked with (*) correspond to Raman modes arising from $N_2$ and $O_2$ solid phases, which were unintentionally introduced as minor phase in Helium transmitting medium in the DAC preparation procedure. For the $PrMnO_3$ and $NdMnO_3$ cases, the Raman spectrum of the background is also shown for comparison.

The evolution of the Raman spectra for all compounds share an overall common behavior: the Raman bands shift towards higher wave numbers due to an overall pressure-induced bond shortening and volume reduction, and become broader. The most notable result of Figure 3 is the complete extinction of the Raman signature of all manganites above a certain pressure, whose value strongly depends on the rare-earth cation. Following similar observations in, e.g. $LaMnO_3$, $GdMnO_3$ or $BiMnO_3$ [5,10,15,27], we interpret this phenomenon as an insulator-to-metal transition and define the critical pressure $P_{IM}$ as the pressure where the Raman signal measured upon increasing pressure could no longer be detected. The transition is reversible and the Raman spectrum is recovered upon decreasing pressure with a significant hysteresis ranging between 3 to 6 GPa depending on the rare-earth cation. Apart from this transition, no anomaly is detected in the spectra; the characteristic orthorhombic spectral profile is maintained up to the disappearance of the Raman signal, indicating that no change of symmetry occurs in the insulating phase.

Figure 4 depicts the pressure dependence of the Raman wavenumbers determined from the fits to the spectra. This pressure evolution does not show any abrupt change until the transition pressure. We note that two pairs of modes at around 473/487 $cm^{-1}$ and 502/522 $cm^{-1}$ cross upon increasing pressure for *A*$MnO_3$ compound with *A* = Sm to Dy. This crossing is allowed by symmetry because these modes corresponds to $A_g(1)$ or $A_g(3)$ mixed modes that cross $B_{2g}(2)$ and $B_{2g}(3)$ modes, respectively, and does not imply any phase transition or change in symmetry.

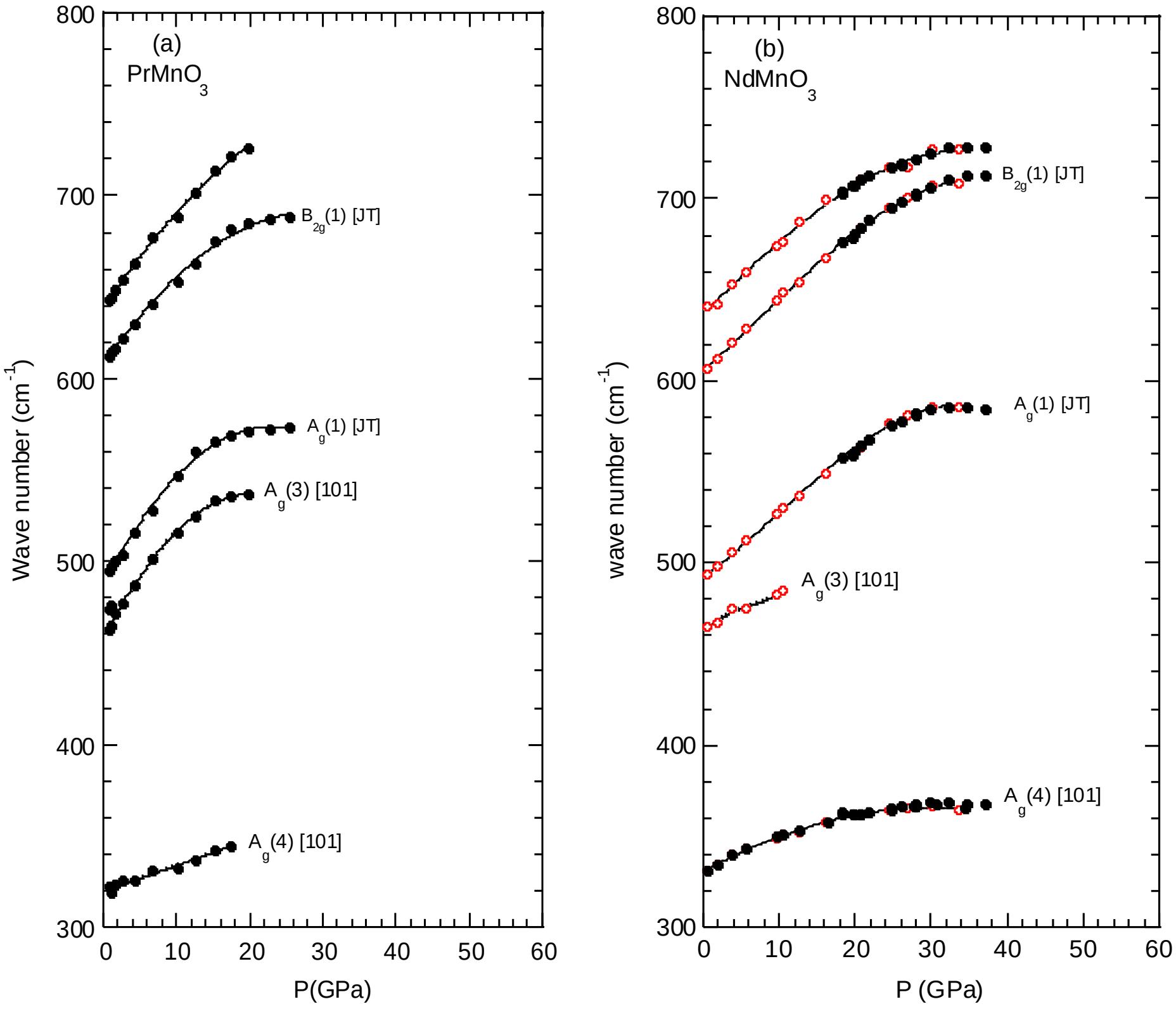

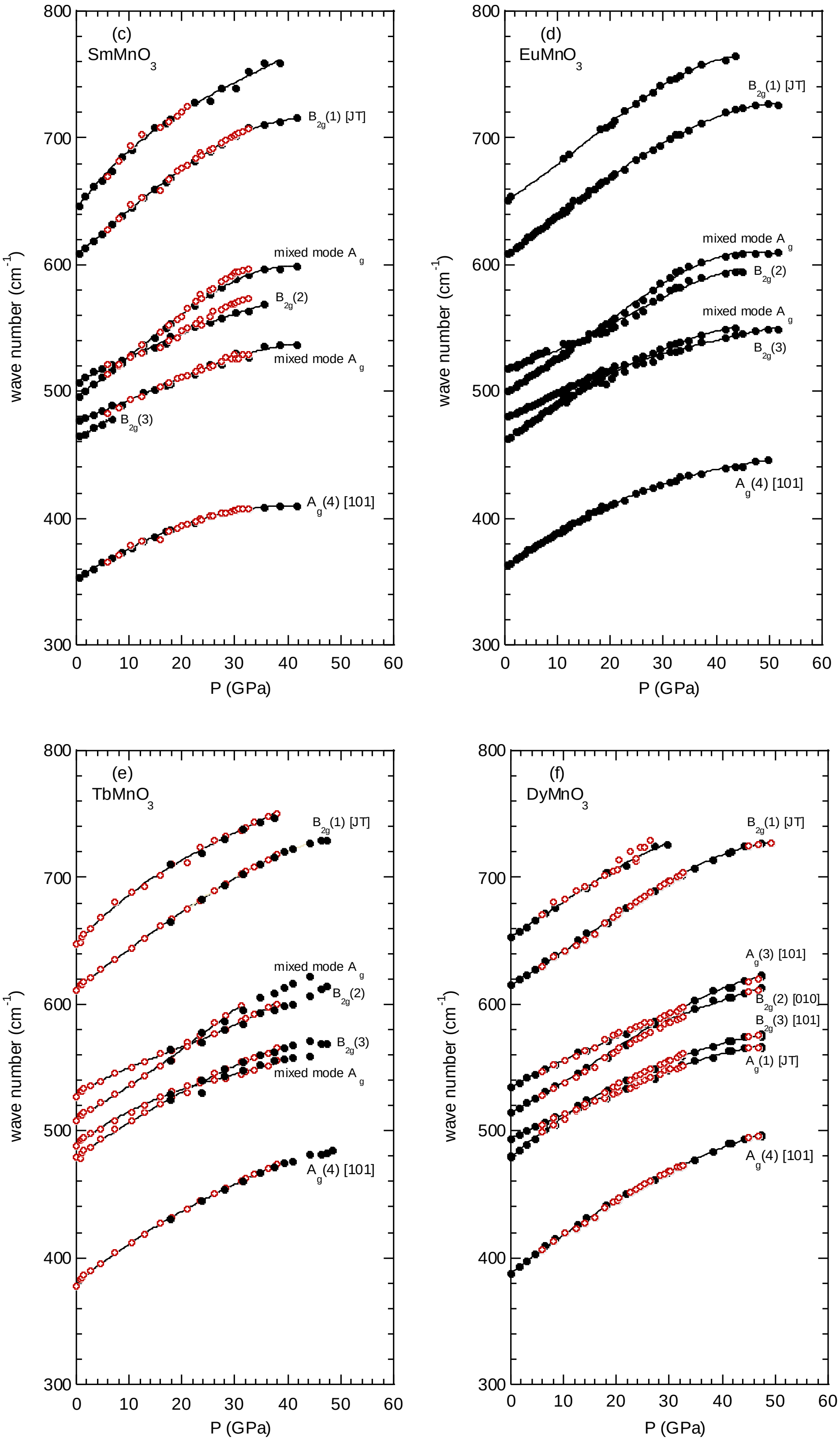

**Figure 4**. Wave number of observed Raman bands as a function of pressure, and mode assignment according to [21]. The basic distortion responsible for the mode activation is also presented. $A_g(4)$: out-of-phase $MnO_6$ [101] rotation, $B_{2g}(1)$: in-plane O2 stretching Jahn-Teller. Black circles: increasing pressure; open red circles: decreasing pressure. The solid lines are guides for the eyes.

## c. High pressure x-ray diffraction

We have performed x-ray diffraction experiments in order to check whether or not the phase transition identified in Raman spectroscopy is accompanied by a symmetry breaking transition, and to investigate the compression mechanisms. Figure 5 shows representative powder x-ray diffraction patterns of *A*$MnO_3$ compounds, with *A* = Pr, Sm, Eu, Dy, recorded at different pressures.

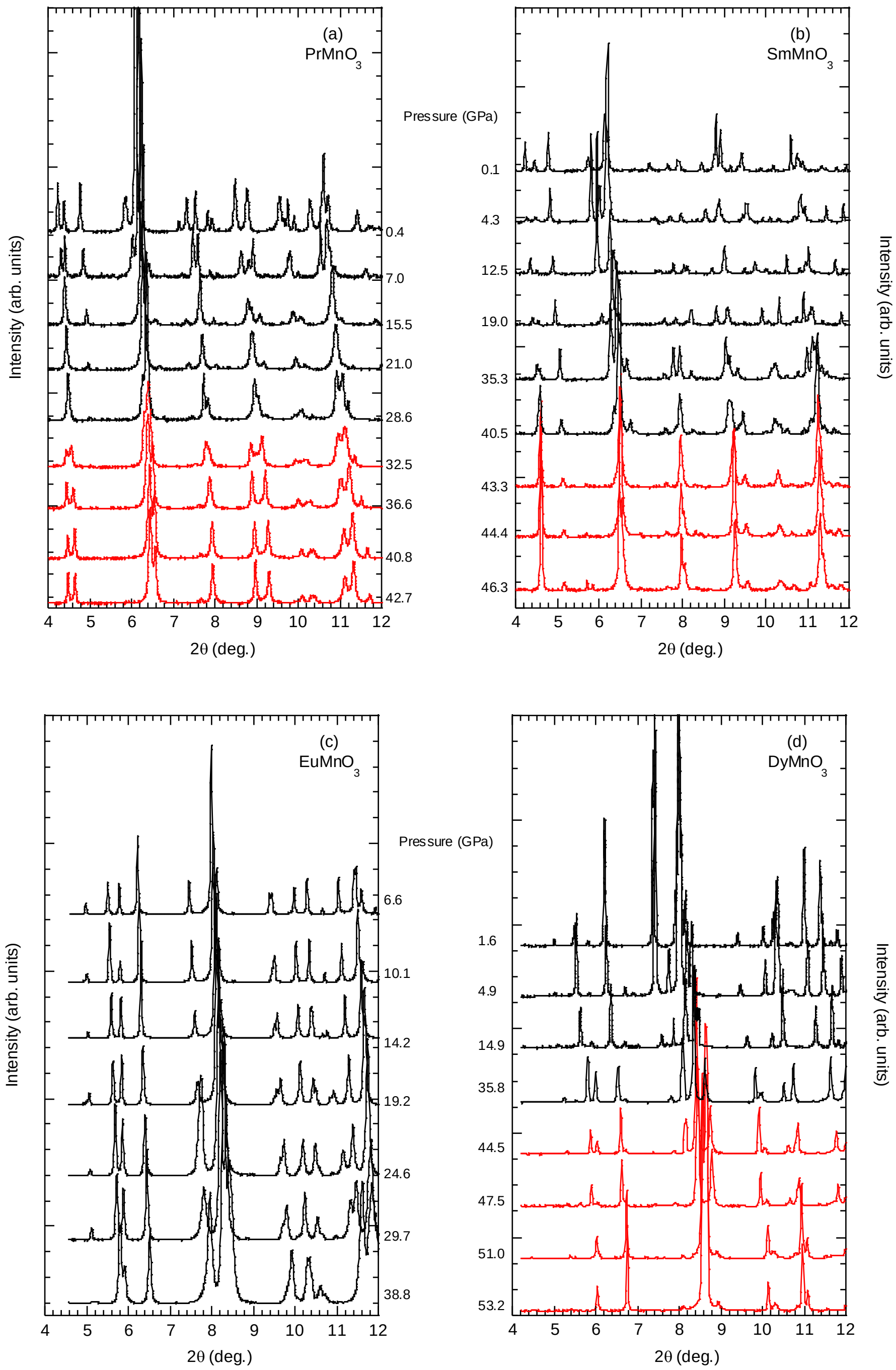

**Figure 5**. X-ray diffraction patterns of (a) $PrMnO_3$ and (b) $SmMnO_3$, recorded at PETRA III (λ = 0.29118 Å), (c) $EuMnO_3$, and (d) $DyMnO_3$, recorded at ID27-ESRF (λ = 0.3738 Å) at selected pressures. The transition to a new phase is marked by a change of color.

The inspection of the XRD spectra presented in Figure 5 shows that for all studied compounds, the patterns maintain the characteristic P*nma* profile up to a critical pressure that matches the transition pressure $P_{IM}$ identified in the Raman evolutions (except for $EuMnO_3$ for which we could not reach $P_{IM}$ in the XRD experiment). Above $P_{IM}$, a clear change of the XRD profile is observed, revealing a structural phase transition. In all cases, we observe a coexistence of phases and a hysteretic behavior upon decreasing pressure, which indicates the first-order nature of the transition.

Possible structures for high-pressure phases were tested by profile matching. For $PrMnO_3$, a good profile matching was obtained for a tetragonal *I4/mcm* cell with $a \approx a_{pc}\sqrt{2}$ and $c \approx 2a_{pc}$. This space group is very common in perovskites and corresponds to the $a^0a^0c^-$ tilt system. The other manganites gave unfortunately less convincing results due to the limited quality of the diffraction diagrams and the width of diffraction peaks, most Bragg reflections being broader than in the low-pressure phase. For $SmMnO_3$, the same tetragonal space group *I4/mcm* (a=5.06 Å, c=7.27 Å at 43.2 GPa) gives somewhat reasonable results. For $DyMnO_3$, the profile matching is still compatible with the P*nma* symmetry, and all other trials gave unreasonable profiles.

In figure 6, we show the pressure dependence of the pseudocubic lattice parameters defined as $a_{pc} = a/\sqrt{2}$, $b_{pc} = b/2$ and $c_{pc} = c/\sqrt{2}$. For $PrMnO_3$ (resp. $DyMnO_3$), the lattice parameters above $P_{IM}$ are determined under the assumption of a tetragonal I4*/mcm* phase (resp. orthorhombic P*nma*). For $SmMnO_3$, whose high-pressure structure could not be determined reliably, we do not show lattice parameters in the high-pressure phase. The quality of the data precluded meaningful Rietveld refinements, so that a detailed analysis of the relevant bond lengths and angles was not possible.

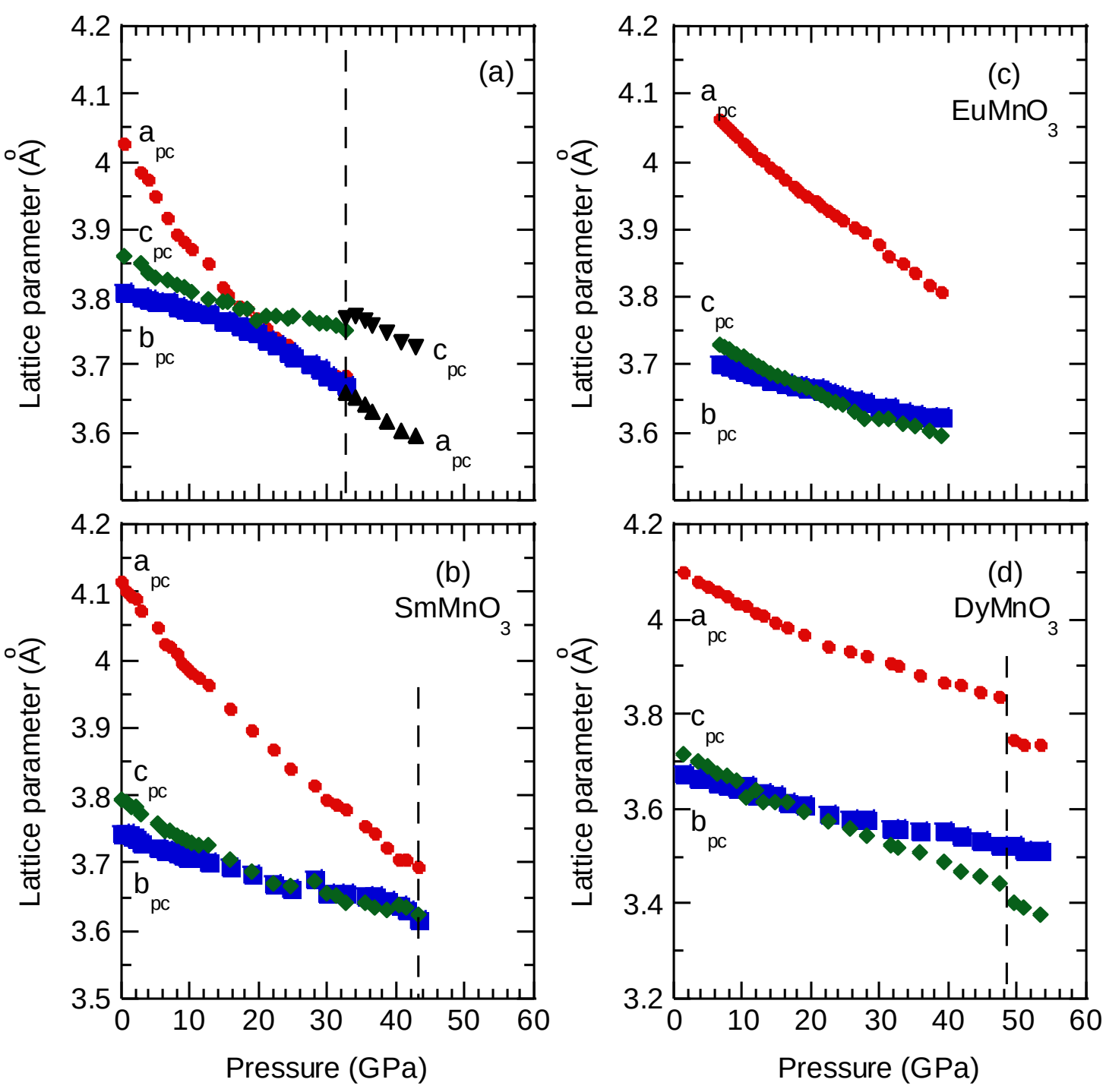

**Figure 6**. Pressure dependence of the pseudocubic lattice parameters of $PrMnO_3$, $SmMnO_3$, $EuMnO_3$ and $DyMnO_3$: red circles $a_{pc}$, green diamonds $c_{pc}$, blue squares $b_{pc}$. The vertical dashed lines mark the critical pressure $P_{IM}$.

Figure 7 shows the pseudocubic volume as a function of pressure of the studied compounds. In the orthorhombic phase up to $P_{IM}$, the pressure dependence of the unit cell volume of all studied compounds could be modelled adequately by a Murnaghan equation of state [28]:

$$V(P) = V_o\left[1 + B_o'\frac{P}{B_o}\right]^{-1/B_o'} \quad (1)$$

where $V$ is the pseudo-cubic volume at the pressure $P$, defined as $V=abc/4$, $V_0$ its value at room pressure, $B_0$ the bulk modulus and $B'_0$ its pressure derivative, all taken at room pressure. Figure 7(d) shows a 4% volume drop at the transition in $DyMnO_3$ associated to the first-order character of the transition. In $PrMnO_3$, such a drop is not observed, nevertheless because we observed a clear phase coexistence and because there is no group/subgroup relation between the high pressure phase and the low-pressure phase, we can argue that the transition is weakly first-order.

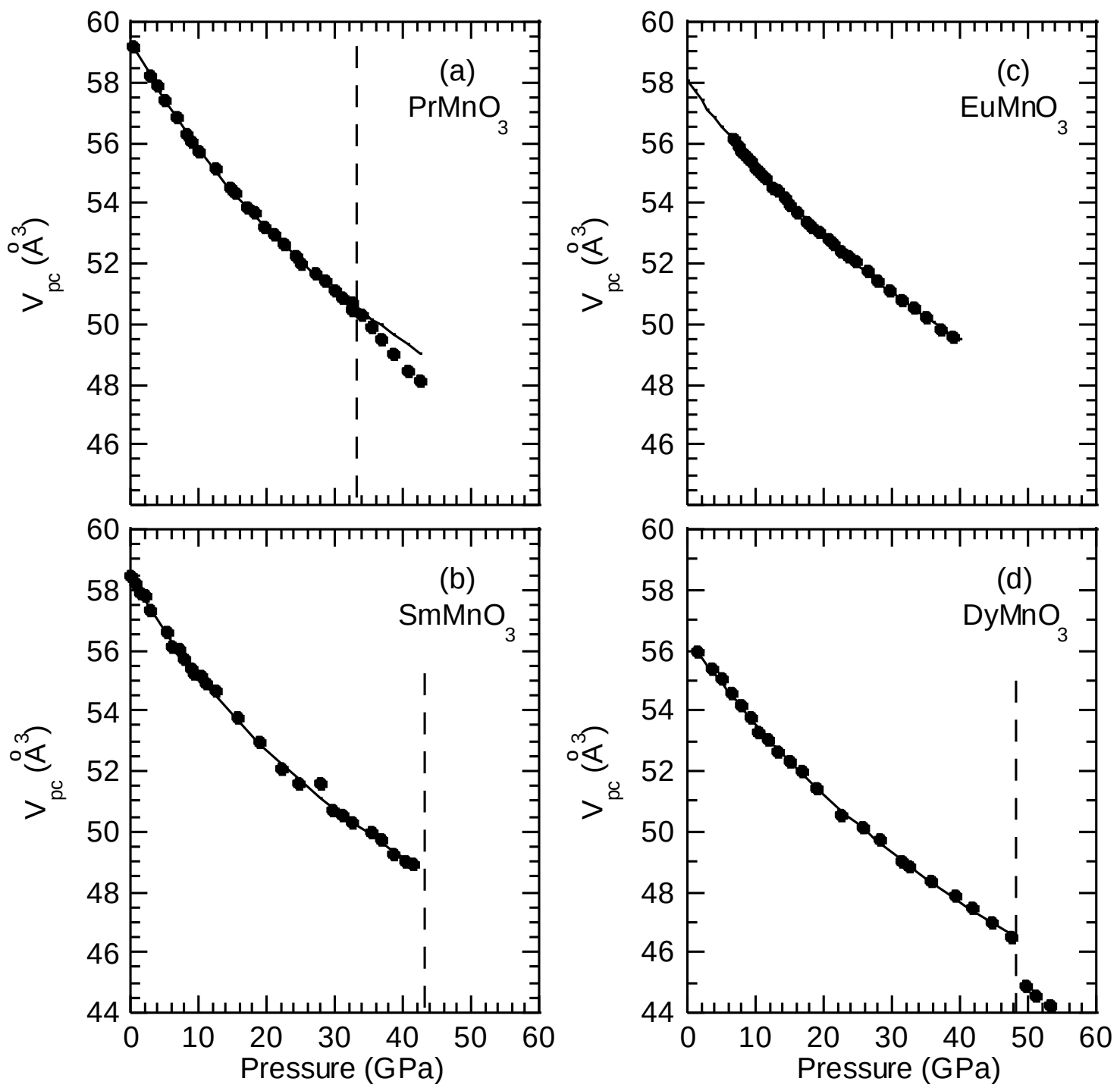

**Figure 7**. Pressure dependence of the pseudocubic volume of $PrMnO_3$, $SmMnO_3$, $EuMnO_3$ and $DyMnO_3$. The solid lines are the best fit to the Murnaghan equation of state (Equation 1). The vertical dashed lines mark the critical pressure $P_{IM}$.

Table 1 presents the values of $B_0$ and $B'_0$, obtained from the fit. The bulk modulus increases as the *A*-cation size decreases from La to Eu and then remains constant for Eu-Dy. We note that the increase of the bulk modulus with decreasing *A*-cation size has been also found in rare-earth chromites $SmCrO_3$, $EuCrO_3$ and $GdCrO_3$ [29]. Moreover, the pressure derivative of the bulk modulus decreases with decreasing the A-cation size, up to $GdMnO_3$, where the minimum value is reached. It is worth to stress that the $B'_0$ value for $LaMnO_3$ is unusually high, far above the values obtained for the other rare-earth manganites or similar perovskites [5].

**Table 1**. Values of the bulk modulus $B_0$ and its pressure derivative $B'_0$, obtained from the best fit of the equation of state to the pseudocubic volume.

| | $B_0$ (GPa) | $B'_0$ | Ref. |
|---|---|---|---|
| $LaMnO_3$ | 97 ± 2 | 9.7 ± 0.4 | [5] |
| $PrMnO_3$ | 139 ± 4 | 4.5 ± 0.3 | This work |
| $SmMnO_3$ | 149 ± 6 | 4.6 ± 0.1 | This work |
| $EuMnO_3$ | 170 ± 5 | 4.4 ± 0.2 | This work |
| $GdMnO_3$ | 170 ± 1 | 3.5 ± 0.07 | [15] |
| $DyMnO_3$ | 170 ± 5 | 3.7 ± 0.2 | This work |

## 4. Discussion

### a. High pressure phase transitions.

For all $A$MnO$_3$ compounds with $A$ = Pr, Sm, Eu, Dy, the Raman and XRD results show the existence of a first-order phase transition occurring at high pressure and characterized by the vanishing of the Raman spectrum and a structural transition.

The interpretation of this transition as an insulator-to-metal transition follows the reasoning already given for $GdMnO_3$ or $BiMnO_3$ [15,27] and that we just recall briefly here. Generally speaking, the vanishing of the Raman spectrum in a perovskite can be accounted for in two ways: 1) by a transition to the cubic phase where no Raman mode is allowed by symmetry or 2) by a transition to a metallic state where the Raman signal is extremely weak due to the strong reduction of the penetration depth of the laser. Since our XRD results show that no compound undergoes a transition to the cubic phase, we discard the first option and conclude that all compounds exhibit an insulator-to-metal phase transition.

Figure 8 shows the transition pressures determined from Raman and XRD data for all compounds. Estimation of the critical pressure from the XRD data is in general more accurate than from the Raman data, since it is revealed by clear changes in the diagrams and abrupt changes in the volume or lattice parameters. Determination of the transition pressure from the Raman spectra on the other hand requires determining the point where the spectrum vanishes, with is sometimes unclear when the signal is already weak with broad features. This could explain the large difference between the two techniques for $GdMnO_3$ [15]. The transition pressure tends to increase for smaller cations, with a possible stabilization at approximately 50 GPa for Eu-Dy.

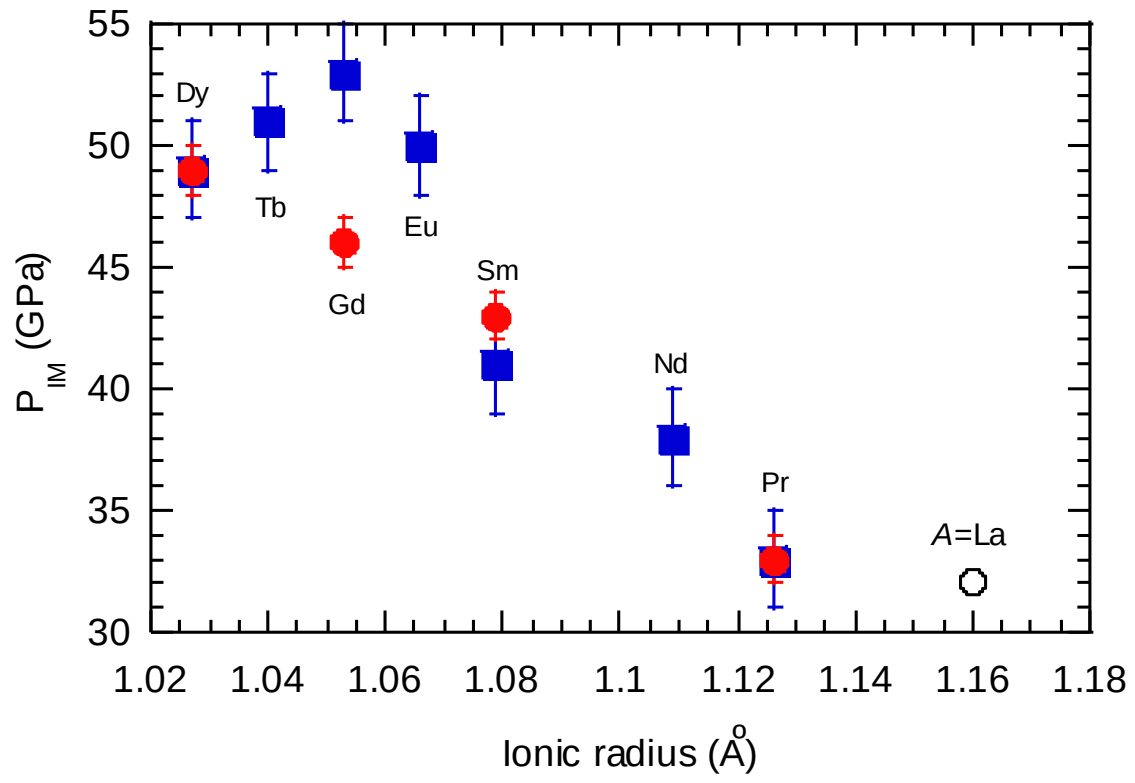

**Figure 8**. Critical pressures as a function of ionic radius determined from XRD (red circles) and Raman spectra (bleu squares). The value of $P_{IM}$ of $LaMnO_3$ and $GdMnO_3$ were obtained from Refs. 5 and 15, respectively.

Unfortunately, the information regarding the crystal structures above the transition remains incomplete and unclear and will require future work. However, we note that, remarkably, no common rule seems to emerge from these preliminary results. In fact, the high pressure phases proposed so far are very different and even belong to different crystal systems ($GdMnO_3$: cubic $P2_13$ [15], $PrMnO_3$: tetragonal *I4/mcm*; $DyMnO_3$: orthorhombic *Pnma*). To this diversity, we can add the case of $BiMnO_3$ that exhibits an intermediate monoclinic polar phase separating the insulating *Pnma* phase and the high-pressure metallic phase [27]. The knowledge of these structures for all compounds would be helpful in order to elucidate the mechanisms for the high-pressure phase transition.

## b. Compression mechanism

We focus here on the compression mechanisms restricting the discussion to the pressure range where the behavior (i.e. Raman frequencies, lattice parameters etc.) can be considered a linear function of pressure. We first discuss the frequencies of the Raman modes and their rate of change measured by their slope. Since different modes are sensitive to different structural aspects, the comparison of their slopes from compound to compound should enable us to compare the different accommodation of pressures in all compounds.

As detailed previously, the out-of-phase $MnO_6$ rotation mode $A_g(4)$ probes the tilting of the $MnO_6$ chain network and the in-plane O2 symmetric stretching mode $B_{2g}(1)$ mirrors Mn-O2 bond length changes. Those modes are therefore excellent sensors of the structural distortions induced by pressure. Furthermore, these bands are well defined and do not overlap with others, so that their frequencies can be determined reliably. Figure 9 shows their slope, and the insert depicts the "mode compressibility" β defined as:

$$\beta = \frac{1}{\omega_0}\left(\frac{d\omega}{dP}\right)_0 \qquad (2)$$

where ω is the mode wavenumber at pressure P and $\omega_0$ and $\left(\frac{d\omega}{dP}\right)_o$ are the mode wavenumber and its slope at room pressure. This parameter measures the relative linear change of the wave number of the aforementioned modes with pressure. We also show the linear compressibility of the lattice parameters defined in a similar way:

$$\delta = -\frac{1}{\ell_0}\frac{d\ell}{dP} \tag{3}$$

$\ell$ being the lattice parameter at pressure P, and $\ell_0$ its value at room pressure.

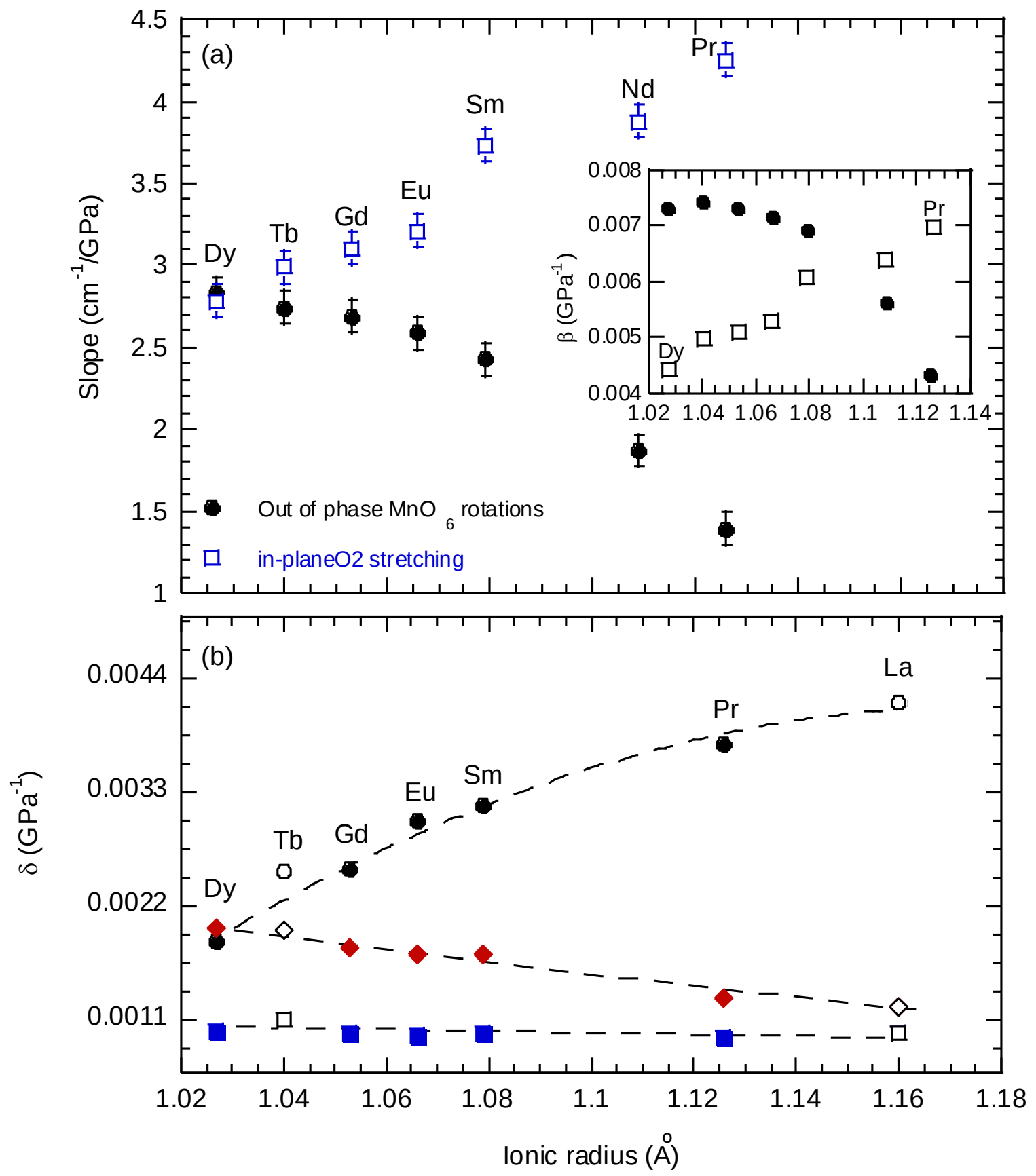

**Figure 9**. (a) Slope of the linear pressure dependence of the wave number of the out-of-phase $MnO_6$ rotations and of the in-plane O2 symmetric stretching modes. The insert depicts the coefficient β, defined by Eq. (2). (b) Compressibility of the lattice parameters as a function of the A-cation size: $\delta_a$ circles, $\delta_b$ squares, and $\delta_c$ diamonds. Closed symbols are our data; open symbols for $LaMnO_3$ and $TbMnO_3$ were calculated from the data reported in Refs. 5 and 12, respectively. The dashed lines are guide for the eyes.

The plot of the linear compressibilities shows that the hard direction is the *b*-axis in all cases: the compressibility along the *b*-axis is the smallest and can virtually be considered independent on the rare-earth ionic radius. In contrast, in the *ac* plane, which contains the two Mn-O2 distances, the manganites exhibit markedly different behavior. For the largest ionic radii, the compression is highly anisotropic, the *a*-axis being softer than the *c*-axis. In parallel, the slope of the in-plane

O2 symmetric stretching mode $B_{2g}(1)$ is much higher than the slope of the out-of-phase $MnO_6$ rotation mode $A_g(4)$. As the ionic radius becomes smaller, the slopes of the two Raman modes converge, as well as the compressibilities along the *a*- and *c*-axis, and become practically equal for $DyMnO_3$. The mode compressibilities β for the two Raman modes follow similar trends and cross between Sm and Pr. Considering β as a measure for the role of the distortion in the volume reduction, this means that for large cations (La, Pr), the change in the tilt angle plays a minor role in the volume reduction under pressure as compared to the reduction of Mn-O2 bonds, with the converse trend for smaller cations.

The frequency of the O2 symmetric stretching mode is in principle sensitive to an average Mn-O2 bond length and therefore does not allow us to follow separately the long (*l*) and short (*s*) Mn-O2 bond lengths, or the difference *l*-*s*, which can be seen as a simplified measure of the Jahn-Teller distortion. On the other hand, the $\delta_a$ and $\delta_c$ compressibilities can reflect this, since the long and short Mn-O2 bonds have their main projections along the *a* and *c*-axes respectively [30]: a larger compressibility of the *a*-axis indicates a stronger reduction of the long Mn-O2 bond length as compared to the short one, and therefore a reduction of the Jahn-Teller distortion. With this in mind, our data across the whole series in figure 10 leads to the conclusion that the reduction of Jahn-Teller distortion under pressure is very rapid for large cations (La-Pr) but gets slower and slower as the ionic radius decreases. For $DyMnO_3$, $\delta_a$ and $\delta_c$ are almost equal, i.e. the compression is almost isotropic in the *ac* plane, implying that the Jahn-Teller distortion hardly changes. This interpretation is in good agreement with the available data on Mn-O2 bond lengths measured by Rietveld refinements for $LaMnO_3$, $TbMnO_3$, $DyMnO_3$, and $GdMnO_3$ [5,12,13,15] that confirm that the reduction of the long Mn-O2 bond (and the difference *l*-*s*) is very pronounced in $GdMnO_3$ and $LaMnO_3$ but extremely slight in $TbMnO_3$ and $DyMnO_3$. Suppression of the Jahn-Teller distortion for the smallest cations, if it occurs at all, can only be expected at very high pressure, and we may even speculate that it persists up to $P_{IM}$.

This interpretation is also consistent with the increase of the critical pressure of the insulator-to-metal transition. Assuming that the delocalization of the electronic states is associated with the disappearance of the Jahn-Teller distortion, we would expect the transition to occur earlier for those compounds in which the Jahn-Teller distortion is reduced more rapidly.

### c. Jahn-Teller distortion: spontaneous strain and Raman spectroscopy

In the following, we will examine the evolution of the Jahn-Teller distortion at high pressure by means of spontaneous strains and Raman spectroscopy. The study of strains can be of high importance to understand the structural behavior of perovskites with cooperative Jahn-Teller distortion. Carpenter *et al* have developed a Landau formalism accounting for the structural evolution of perovskites containing Jahn-Teller cations and octahedral tilting [31,32]. This approach gives a detailed account of the relationships between macroscopic strains and order-parameters, in order to disentangle the combined effects of tilting and Jahn-Teller distortions in perovskites. For the P*nma* perovskites, the total distortion involves three different order parameters. Two of them are associated with the out-of-phase rotations along *a* and c (order parameter $q_4$, linked to the Ag(4) Raman mode) and the in-phase rotation along *b* ($q_2$, linked to $A_g(2)$ Raman mode), while the third one is associated with the Jahn-Teller distortion. Carpenter *et al.* [31,32] have shown that all three order parameters couple with spontaneous strains, the shear strain $e_4$ being the most insightful component to follow for a discussion of Jahn-Teller distortion due to its more simple dependence on the order-parameters.

Before discussing Jahn-Teller distortion, we need to consider the evolution of tilts, which can be made from the evolution of the corresponding Raman bands. For each $AMnO_3$ compound we could follow the $A_g(4)$ and $B_{2g}(1)$ Raman mode however the $A_g(2)$ is much more difficult to follow for several reasons. First because the $A_g(2)$ mode is located at low wavenumber (<300 cm$^{-1}$), second because it can mix with the $A_g(7)$ A-shift mode depending on the A-cation size, and finally because it is hardly seen for large rare-earth for which the orthorhombic distortion is small and consequently the Raman spectrum is broad and weak in intensity. Nonetheless, for $GdMnO_3$ [15], we have measured pressure variations of 2.47 cm$^{-1}$/GPa and 1.06 cm$^{-1}$/GPa for the $A_g(4)$ and the $A_g(2)$ tilt modes respectively. If we consider that these two modes have the same linear dependence with the tilt angles of 23.5 cm$^{-1}$/deg, as demonstrated by Iliev [21], we can expect pressure variations of 0.105 °/GPa and 0.0451 °/GPa for the [101] and the [010] tilt angles respectively. Thus we assume a tilt increase of 4.7° (28%) and 2° (14%) for the [101] and the [010] tilt angles, respectively, for a pressure increase of 45 GPa in $GdMnO_3$. As discussed before, the tilt increase is even lower for large *A*-cation such as $PrMnO_3$ for which we can extract a pressure variations of 0.059°/GPa or a tilt increase of 1.9° (15%) for the [101] tilt angle for a pressure increase of 32 GPa. This simple consideration show that for most of the $AMnO_3$ compounds the tilt angles will increase little with pressure (not greater than 30%) and that a complete suppression of the Jahn-Teller distortion ($q_{2JT}=0$) should be detected as an anomaly in the pressure evolution of the shear strain, as it was claimed in Ref. [32] for $LaMnO_3$ on the basis of the experimental XRD data from Ref. [5].

In the following, we will carry out a similar analysis for our series of compounds. For consistency, we calculate the shear strain following the definitions by Carpenter. As a reference state, we choose the axes of the simple cubic perovskite cell, with a reference cubic parameter $a_0$ defined as:

$$a_o = \sqrt[3]{V_p} \quad (6)$$

with $V_p$ the pseudocubic volume. The shear strain is defined as:

$$2e_{23} = e_4 = \frac{a_{pc} - c_{pc}}{a_o} \quad (9)$$

where $a_{pc}$, $b_{pc}$ and $c_{pc}$ are the pseudocubic lattice parameters. Figure 10 shows the pressure evolution of the shear strain $e_4$, for all the compounds studied in this work. We have also included the data of $LaMnO_3$ and $GdMnO_3$, obtained from the analysis of data reported in [5,15].

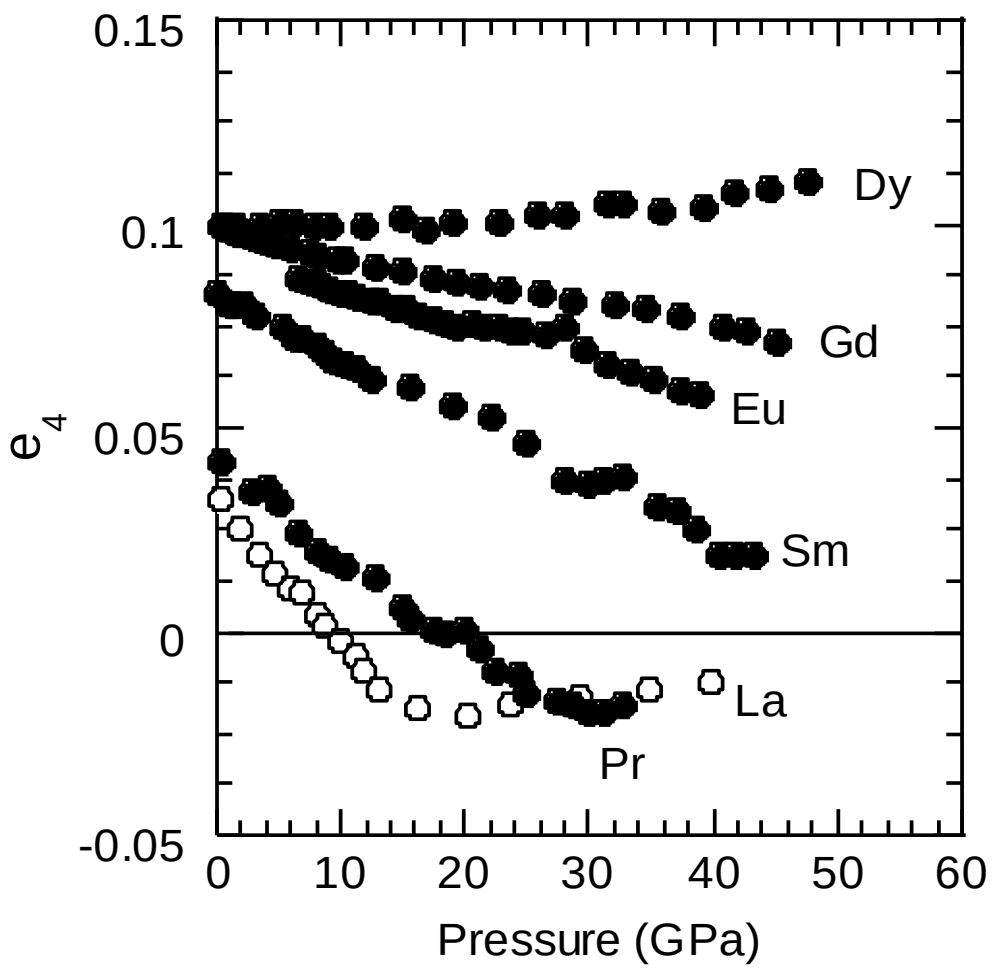

**Figure 10**. Pressure dependence of the shear strain $e_4$ for $R$MnO$_3$, with $R$ = La, Pr, Sm, Eu, Gd, Tb and Dy. Data for $LaMnO_3$ and $GdMnO_3$ were taken from Refs. 5 and 15, respectively.

The shear strain $e_4$ is always positive at room pressure. Only for $LaMnO_3$ and $PrMnO_3$ does the shear strain exhibit an anomaly at around 16 GPa and 30 GPa, respectively. For $SmMnO_3$, $EuMnO_3$ and $GdMnO_3$, $e_4$ decreases with pressure but does not show any anomaly and remains always positive. Therefore, for $SmMnO_3$, $EuMnO_3$, $GdMnO_3$ and $DyMnO_3$, the strain analysis does not give any evidence for a suppression of the cooperative Jahn-Teller distortion under pressure.

The trends for the evolution of Jahn-Teller under pressure call for comment in the light of some qualitative statements that can be made based on the lattice parameters only. At ambient pressure, all orthomanganites exhibit $a_{pc} > c_{pc} > b_{pc}$, which is characteristic of *Pnma* perovskites having both octahedra tilting and Jahn-Teller distortion, while $a_{pc} > b_{pc} > c_{pc}$ is usually typical of pure tilt systems such as the closely related orthoferrites $A$FeO$_3$ [30], and $c_{pc} > a_{pc} > b_{pc}$ is found for the *Pnma* perovskites with the largest *A*-cations, low tilt angles and low strains that are dominated by intrinsic shear strain of the octahedra [33,34]. Under pressure, we see (figure 6) that for $EuMnO_3$, $GdMnO_3$ and $DyMnO_3$, $b_{pc}$ and $c_{pc}$ cross so that they exhibit $a_{pc} > b_{pc} > c_{pc}$ although Jahn-Teller distortion is still present, showing that this lattice-parameter crossing alone is not in general an indication for a reduction of Jahn-Teller distortion.

The average information given by XRD can now be compared to the Raman spectroscopy results. As a local probe, Raman spectroscopy is often more sensitive to small structural rearrangements. In previous studies, the characteristic signature of the reduction of Jahn-Teller effect in $LaMnO_3$ and related compounds was the splitting of the $B_{2g}$ stretching mode at high frequency and a gradual intensity transfer from one component to another [5,10]. This intensity transfer was assigned to the formation of “domains” of distorted and regular octahedra, observed from 3 to 34 GPa, and the gradual disappearance of the former. Here, our spectra at ambient conditions (figure 1) do show a small peak appearing as a shoulder of the $B_{2g}$ (1) mode, and it can be followed with pressure (figure 4), but we do not observe any intensity transfer for any of the compounds, including $PrMnO_3$. This was also the case in $BiMnO_3$ [27]. This suggests that in our experiments, if coexistence of distorted and undistorted octahedral actually takes place, it is not significantly affected by pressure. Since diffraction gives nonetheless a clear evidence for a reduction of Jahn-Teller distortion, at least at low pressures, this can be interpreted by a homogeneous reduction of the Jahn-Teller distortion throughout the sample. This point, however, needs to be clarified by further studies.

### d. Hydrostatic pressure versus chemical pressure

We know discuss our high-pressure data in the light of the concept of "chemical pressure" in rare-earth manganites. Generally speaking, in a solid solution, the substitution of a rare-earth on the A site by a smaller isovalent cation causes a reduction of cell volume and is therefore depicted as a positive (compressive) chemical pressure. Because the substituted cations are supposed to be randomly distributed, this chemical pressure is assumed isotropic; i.e. implicitly compared to hydrostatic pressure. For the case of rare-earth manganites, we know indeed that the volume depends roughly linearly on the ionic radius (see Figure 2(d)), and the lattice parameters can be expected to follow Vegard's law [35]. Would the analogy between hydrostatic and chemical pressures be perfect, then the lattice parameters and other physical properties of a given orthomanganite should just depend on its volume, irrespective of whether this volume is reached by chemical substitution or hydrostatic pressure. Our experimental data enable us to examine the validity of this analogy by plotting physical quantities as a function of volume across the full $AMnO_3$ series.

Figure 11 shows the pseudocubic lattice parameters of orthorhombic rare-earth manganites as a function of the cubic parameter $a_0=V^{1/3}$ for different rare-earth manganites at room conditions [22-25,36]. Under the assumption of Vegard's law, the lattice parameters of a given solid solution under the effect of "chemical pressure" are expected to lie on a straight line between the two end compounds. Figure 11 shows that the chemical pressure does not act in the same way in the three lattice parameters of the rare-earth manganites. While pseudocubic lattice parameters $b_{pc}$ and $c_{pc}$ decrease monotonously with decreasing ionic radius from $La^{3+}$ to $Lu^{3+}$, $a_{pc}$ shows a non-monotonous behavior, with a maximum located around Eu. On the same plot we show a selection of data measured under hydrostatic pressures. For the smallest A-cations (say from Dy to Lu), the hydrostatic pressure evolution does more or less follow the line of cation substitution. For the largest A cations, however, a discrepancy is immediately apparent: the lattice parameters under hydrostatic pressure for $LaMnO_3$ do not at all follow the La-Pr line, the most striking case being $a_{pc}$ which decreases under hydrostatic pressure whereas it increases under chemical pressure. The analogy between hydrostatic pressure and chemical pressure therefore appears reasonable for small cations but misleading for larger cations. Similar conclusions could be drawn by plotting e.g. the Raman frequencies as a function of volume.

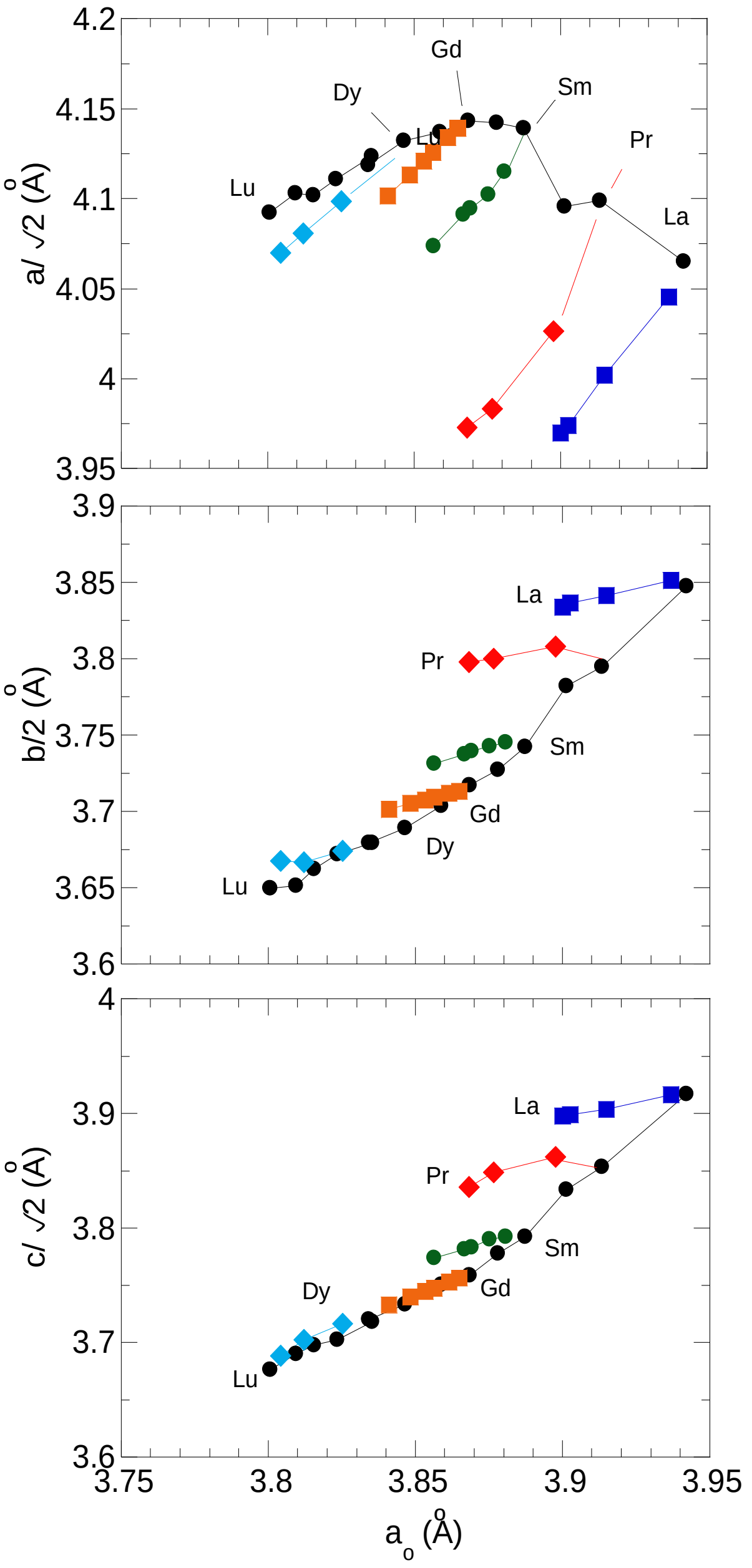

**Figure 11**. Pseudocubic lattice parameters of some orthorhombic rare-earth manganites, as a function of $a_0 = V^{1/3}$, measured at room conditions and at different fixed hydrostatic pressures.

This difference between chemical pressure and hydrostatic pressure merely reflects the diversity and complexity of the compression mechanisms at work in the manganites. The analogy is valid when compression is isotropic and realized through a homogeneous shortening of all chemical bonds, which is approximately realized in manganites with small A cations (≈ Gd-Dy). Conversely, the analogy fails for manganites with larger cations because other compression mechanisms come into play (Jahn-Teller distortion and tilts).

## 5. Conclusions

In this work we report a high pressure Raman scattering study in orthorhombic rare-earth manganites $AMnO_3$, with $A$ = Pr, Nd, Sm, Eu, Tb and Dy, and synchrotron X-ray diffraction in $A$ = Pr, Sm, Eu, and Dy. Our data, together with a review of previous studies, provide an overview on the behavior of manganites under hydrostatic pressure. The general picture shows a gradual evolution across the rare-earth series between two different behaviors.

For small cations, the compression under hydrostatic pressure is close to be isotropic in the *ac* plane, with only small evolutions of the tilt angles and cooperative Jahn-Teller distortion, as revealed by the evolution of the shear strain. The bulk modulus (≈ 170 GPa) and the transition pressure (≈ 48 GPa) are nearly independent of the cation and the effect of chemical pressure and hydrostatic pressure are very similar. On the other hand, for large cations, the reduction of the cooperative Jahn-Teller distortion contributes significantly to the volume reduction. The bulk modulus and the transition pressure decrease strongly as the ionic radius increases. The anomaly of the shear spontaneous strain suggests a suppression of the average cooperative Jahn-Teller distortion before the phase transition is reached. Chemical substitution and hydrostatic pressure have distinctly different effects.

In all cases, there is evidence for a structural and insulator-to-metal phase transition at high pressure. The crystal structures of the high-pressure phases to vary from compound to compound and no general rule emerges so far. This variety however can be understood if the different compression behaviors described above lead to different mechanisms at the phase transition. Yet, several aspects call for future work: the detailed crystal structures of the high-pressure phases, the possibly different mechanisms of the insulator-to-metal transition for the different rare-earths, the process by which Jahn-Teller distortion is reduced by pressure (with or without coexistence of distorted and undistorted octahedral) and the associated Raman signature. Further studies are therefore needed for a thorough understanding of the behavior of rare-earth manganites under hydrostatic pressure.

**Acknowledgements**

This work was supported by Fundação para a Ciência e Tecnologia, through the Project PTDC/FIS-NAN/0533/2012 and by QREN, through Projecto Norte-070124-FEDER-000070 Nanomateriais Multifuncionais. We acknowledge the Deutsches Elektronen Synchrotron, DESY (Petra III), Hamburg, Germany and the European Radiation Facility, ESRF, Grenoble, France for providing beam time. We acknowledge Zuzana Konopkova (DESY), Hanns-Peter Liermann (DESY) and Gaston Garbarino (ESRF) for their support during the experiments.